\documentclass[aps,twocolumn,pra,showpacs,floatfix]{revtex4}
\usepackage{dcolumn}

\begin{document}
\title{All-order calculations of the spectra of Ba~II, Ra~II, Fr~I and
  superheavy elements E119~I and E120~II.}

\author{V. A. Dzuba}

\affiliation{School of Physics, University of New South Wales,
Sydney, NSW 2052, Australia}

\date{ \today }

\begin{abstract}
A technique based on summation of dominating classes of
correlation diagrams to all orders in Coulomb interaction is used to
calculate the energies of the lowest $s,p$ and $d$ states of Ba II, Ra
II, Fr I, E119~I and E120~II. Breit and quantum electrodynamic
corrections are also included. Comparison with experiment  for Ba~II, Ra~II and
Fr~I demonstrates that the accuracy of the calculations is on the level of
0.1\%. The technique has been applied to predict the spectra of superheavy
elements E119 and E120$^+$. The role of the {\em ladder
  diagrams} (V. A. Dzuba, Phys. Rev. A {\bf 78}, 042502 (2008)) which
is the most recent addition to the method has been emphasised. Their inclusion significantly
improves the accuracy of the calculations and expands the
applicability of the method. 

\end{abstract}
\maketitle

\section{Introduction}

Accurate atomic calculations are very important for a number of
applications, such as search for new physics beyond standard model in
the measurements of the parity and time invariance violation in
atoms~\cite{Ginges,PNC}, search for space-time variation of fundamental
constants~\cite{alpha}, prediction of the properties of atoms and ions
were experimental data is poor or absent, etc. The latter include
in particular highly charge ions~\cite{HCI} and superheavy elements
($Z>100$)~\cite{Pershina1,Pershina2}. 

Atoms with one valence electron above closed shells play a special
role in these studies. Their relatively simple electron structure
allows high accuracy of the calculations leading to very accurate
interpretation of experimental data. For example, current best
low-energy test of the standard model is based on the measurements of
parity non-conservation in cesium~\cite{Wood} which has one valence
electron above closed-shells of Xe-like core. Interpretation of the
measurements is based on accurate calculations of Ref.~\cite{Cs-PNC}.
Further progress for the PNC
measurements is considered for atoms and ions which have electron
structure similar to those of cesium. This includes Rb~\cite{Rb},
Ba$^+$~\cite{Ba+}, Ra$^+$~\cite{Ra+}, Fr~\cite{FrPNC}, and Fr-like
ions~\cite{Fr-like}. Accurate predictions of the spectra and other
properties of the superheavy elements were done for E119, and
E120$^+$~\cite{Lim,Eliav,Dinh1,Dinh2,Borsch} which are also analogous
of cesium.  

The most popular method currently used for accurate calculations for atoms
with one valence electron is the coupled cluster (CC) method. Its
accuracy depends on the number of terms included in the expansion of
the wave function. Its simplest version, which includes only terms
with single and double excitations (SD) from the reference ground
state configuration, gives good accuracy only for limited number of
systems. In particular, the accuracy is poor for cesium, and missed
third-order diagrams need to be included for better
accuracy~\cite{Cs-SD}. The accuracy is significantly better if also
valence triple excitations are included (the CCSDvT
approximation)~\cite{Derevianko}. However, the method becomes very
demanding for computer power.

An alternative approach was developed in our group about 25 years
ago. It is based on summations of selected classes of higher-order
correlation diagrams to all orders in Coulomb
interactions~\cite{all-order}. Further in the paper we will call it
the correlation potential method (CPM) for convenience of references.
The following classes of the correlations were included in CPM in all
orders: (a) screening of Coulomb interaction of 
the valence electron with electrons in the core by other core
electrons, (b) interaction between an electron excited from atomic
core with the hole in the core caused by this excitation, and (c)
iterations of the correlation operator $\hat \Sigma$. This turned to
be very powerful method which gives a fraction of a per cent accuracy for the
energies of $s$ and $p$ states of alkali atoms. The computer power
needed for the calculations is small even compared to simple SD
approximation. On the other hand, accuracy for $d$ states and accuracy
for atoms other than alkali was not high. For example, the accuracy
for Ba$^+$ is almost the same as for Cs. This is contrary to what is
expected since the relative value of the correlation corrections are
smaller for Ba$^+$ than for Cs due to two times stronger central
potential. The reason for lower accuracy was explained in
Ref.~\cite{ladder} for the case of cesium and thallium. The fact of
high accuracy for $s$ and $p$ states of alkali atoms and lower
accuracy for other atomic systems is related to a particular choice of
higher-order diagrams included in CPM method. The three classes of
higher-order diagrams listed above dominate in systems where valence
electron is on large distances from the core. When valence electron is
closer to the core another class of higher-order diagrams becomes
important. These diagrams describe residual Coulomb interaction of the
valence electron with the core. When valence electron is close to the
core this interaction becomes strong and needs to be included in all
orders. In the CPM method this interaction is included in second order
only. This limits the accuracy of the calculations for systems where
external electron is close to the core. The most pronounced example is
probably thallium atom where external $6p$ electron
is very close to the $6s$ electrons and strongly interacts with
them~\cite{ladder}. 

A method to include the core-valence residual interaction to
all-orders was developed in Ref.~\cite{ladder}. It is based on
iterations of the equations similar to the CC SD equations. The terms
arising from the iterations of theses equations were described by
diagrams which were called {\em ladder diagrams}. When the
contribution of the ladder diagrams were added to the result of the
CPM calculations it lead to significant improvement of the accuracy of
the calculations for $s, \ p$ and $d$ states of Cs and Tl.

In present paper we further extend the application of the method by
performing calculations for Ba$^+$ and Ra$^+$ ions and demonstrating
that inclusion of ladder diagrams leads to very accurate results. Then
we apply the method to calculate the spectra of superheavy elements
E119 and E120$^+$. Calculations for francium are also included as
another test of the method which helps to estimate the accuracy for
the E119 superheavy element. Breit and QED corrections are included
for higher accuracy. The spectra of E119 and E120$^+$ were considered
before~\cite{Eliav,Dinh1}. However, present paper presents more
complete and accurate results.

\section{Method of Calculations}

Accurate calculations for heavy and superheavy many-electron atoms
need accurate treatment of correlations and relativistic effects. 
We use the all-order correlation potential method~\cite{all-order}
supplemented by inclusion of ladder diagrams~\cite{ladder} to include
dominating correlation effects to all orders in residual Coulomb
interaction. For accurate treatment of relativistic effects we start
the calculations with the relativistic Hartree-Fock (RHF) method based
on solving Dirac-like equations and then we include Breit interaction
and quantum electrodynamic (QED) corrections.

\subsection{Correlations}

Calculations start from the relativistic
Hartree-Fock method (RHF) in the $V^{N-1}$ approximation. States of
valence electron are calculated with the use of the correlation
potential $\hat \Sigma$:
\begin{equation}
(\hat H_0 +\hat \Sigma - \epsilon_v)\psi_v =0.
\label{Brueck}
\end{equation}
Here $\hat H_0$ is the RHF Hamiltonian, $\psi_v$ and $\epsilon_v$
are the wave function and removal energy of the valence electron. 
Correlation potential $\hat \Sigma$ is defined in such a way that its
average value over the wave function of valence electron in state $v$
is the correlation correction to the energy of this state 
\begin{equation}
  \delta \epsilon_v = \int \psi_v(r_1) \Sigma(r_1,r_2) \psi_v(r_2) dr_1dr_2.
\label{def-sigma}
\end{equation}
$\hat \Sigma$ is a non-local operator similar to Hartree-Fock exchange
potential. Many-body perturbation theory expansion for $\hat \Sigma$
starts from second order. The second-order correlation operator $\hat
\Sigma^{(2)}$ has been described in our previous
works~\cite{CPM,ladder}. For most of atomic systems inclusion of just
second-order $\hat \Sigma$ leads to significant improvements of the
accuracy of the calculations. Further improvement is achieved when
higher-order correlations are also included.
Beyond second order we include four
dominating classes of higher order correlations: (a) screening of
Coulomb interaction, (b) hole-particle interaction, (c) iterations of
$\hat \Sigma$, and (d) ladder diagrams. All these higher-order
correlations are included in all orders of residual Coulomb
interaction. 

Two of theses classes of higher-order correlations are included in the
calculations of $\hat \Sigma$~\cite{all-order}, the screening of
Coulomb interaction between valence and core electrons by other core
electrons, and hole-particle interaction between a hole left in the
core by electron excitation and the excited electron. Third chain of
all-order diagrams, the iteration of the correlation potential, is
included by iterating the equations (\ref{Brueck}). Note that the
single-electron wave functions for the states of valence electron
found by solving equations (\ref{Brueck}) are often called Brueckner
orbitals (BO). 

Another chain of all-order diagrams describes residual Coulomb
interaction of external electron with the core. It is included by
solving coupled-cluster-like equations for ladder
diagrams~\cite{ladder}. The equations are obtained by taking
single-double (SD) approximation for the coupled-cluster (CC) method and
removing terms which otherwise would lead to double counting of the
effects which are already included in the correlation potential $\hat
\Sigma$. We stress once more that the most important all-order effect,
the screening of Coulomb interaction, is better treated in the
calculation of the all-order correlation potential $\hat \Sigma$ than
in solving of the CC equations. This is because of the use of the
relativistic Fynman diagram technique while calculating $\hat
\Sigma$. Relativistic technique includes all possible time ordering of
the hole-particle loops which in terms of the CC expansion means
inclusion of selected triple and higher excitations.

The equations for ladder diagrams can be written as two
sets of equations~\cite{ladder}. The first is for atomic core:
\begin{eqnarray}
  && (\epsilon_a+\epsilon_b-\epsilon_m-\epsilon_n)\rho_{mnab} =
  g_{mnab}+ \label{lcore} \\ 
  && \sum_{rs}g_{mnrs}\rho_{rsab}+\sum_{rc}(g_{cnbr}\rho_{mrca}+
  g_{cmar}\rho_{nrcb}) \nonumber .
\end{eqnarray}
And another is for a specific state $v$ of an external electron:
\begin{eqnarray}
  && (\epsilon_v+\epsilon_b-\epsilon_m-\epsilon_n)\rho_{mnvb} =
  g_{mnvb}+ \label{lval} \\ 
  && \sum_{rs}g_{mnrs}\rho_{rsvb}+\sum_{rc}(g_{cnbr}\rho_{mrcv}+
  g_{cmvr}\rho_{nrcb}) \nonumber .
\end{eqnarray}
Here parameters $g$ are Coulomb integrals 
\[ g_{mnab} = \int \int \psi_m^\dagger(r_1) \psi_n^\dagger(r_2)e^2/r_{12}
\psi_a(r_1)\psi_b(r_2)d\mathbf{r}_1d\mathbf{r}_2, \] 
variables $\rho$ are
the coefficients representing expansion of the atomic wave function 
over double excitations from the zero-order Hartree-Fock reference 
wave function; parameters $\epsilon$ are the single-electron Hartree-Fock
energies. Coefficients $\rho$ are to be found by solving the equations
iteratively starting from
\[ \rho_{mnij} = \frac{g_{mnij}}{\epsilon_i+\epsilon_j-\epsilon_m-\epsilon_n}. \]
Indexes $a,b,c$ numerate states in atomic core, indexes
$m,n,r,s$ numerate states above the core, indexes $i,j$ numerate
any states.

The equations for the core (\ref{lcore}) do not depend on the 
valence state $v$ and are iterated first. The convergence is controlled
by the correction to the core energy
\begin{equation}
  \delta E_C = \frac{1}{2}\sum_{abmn} g_{abmn}\tilde\rho_{mnab},
\label{deltaec}
\end{equation}
where
\[ \tilde\rho_{mnab} = \rho_{mnab} - \rho_{mnba}. \]

When iterations for the core are finished the equations
(\ref{lval}) are iterated for as many valence states $v$ as needed.

Correction to the energy of the valence state $v$ arising from the
iterations of equations (\ref{lcore}) and (\ref{lval}) is given by
\begin{equation}
  \delta \epsilon_v = \sum_{mab}g_{abvm}\tilde\rho_{mvab}+
  \sum_{mnb}g_{vbmn}\tilde\rho_{mnvb}.
\label{deltav}
\end{equation}

Since Brueckner energy $\epsilon_v$, in the equation (\ref{Brueck}) 
and the correction $\delta \epsilon_v$, in the equation (\ref{deltav}) 
both include the second-order correlation correction, it is convenient 
to define the correction associated with the ladder diagrams as a difference
\begin{equation}
  \delta \epsilon_v^{(l)} = \delta \epsilon_v - \langle v | \hat \Sigma^{(2)}| v \rangle.
\label{deltal}
\end{equation}
Here $\hat \Sigma^{(2)}$ is the second-order correlation potential.

\subsection{Breit interaction}

We treat Breit interaction in zero energy transfer approximation. The
Breit Hamiltonian includes magnetic interaction between moving
electrons and retardation:
\begin{equation}
\hat H^{B}=-\frac{\mbox{\boldmath$\alpha$}_{1}\cdot \mbox{\boldmath$\alpha$}_{2}+
(\mbox{\boldmath$\alpha$}_{1}\cdot {\bf n})
(\mbox{\boldmath$\alpha$}_{2}\cdot {\bf n})}{2r} \ .
\label{Breit}
\end{equation}  
Here ${\bf r}={\bf n}r$, $r$ is the distance between electrons, and 
$\mbox{\boldmath$\alpha$}$ is the Dirac matrix.

Similar to the way Coulomb interaction is used to form self-consistent
Coulomb potential, Breit interaction is used to form self-consistent
Breit potential. Other words, Breit interaction is included into
self-consistent Hartree-Fock procedure. Thus the important relaxation
effect is included. The resulting inter-electron potential in
(\ref{Brueck}) consist of two terms
\begin{equation}
\hat V=V^{C}+V^{B} \ ,
\end{equation}  
$V^{C}$ is the Coulomb potential, $V^B$ is the Breit potential.
Coulomb interaction in the second-order correlation potential $\hat
\Sigma^{(2)}$  
is also modified to include Breit operator (\ref{Breit}).
The Breit correction to the energy of external electron is found
by comparing the second-order Brueckner energies (Eq.~(\ref{Brueck})) 
calculated with and without Breit interaction.

\subsection{QED corrections}

We use the radiative potential method developed in Ref.~\cite{radpot}
to include quantum radiative corrections. This potential has the form
\begin{equation}
V_{\rm rad}(r)=V_U(r)+V_g(r)+ V_e(r) \ ,
\end{equation}
where $V_U$ is the Uehling potential, $V_g$ is the potential arising from the 
magnetic formfactor, and $V_e$ is the potential arising from the
electric formfactor. The $V_U$ and $V_e$ terms can be considered as
additions to nuclear potential while inclusion of $V_g$ leads to some
modification of the Dirac equation (see Ref.~\cite{radpot} for details).
As for the case of Breit interaction, the QED corrections to the energies
of external electron are found by solving equations (\ref{Brueck}) with
and without radiative potential.

\section{Results and Discussion}

\begin{table*}
\caption{Removal energies (cm$^{-1}$) of the lowest $s,p,d$ states of
  Ba$^+$, Ra$^+$, E120$^+$, Fr and E119 in different approximations
  together with ladder diagram, Breit and QED corrections and
  experimental data. $\Delta = 100 (E_{\rm final} - E_{\rm
    expt})/E_{\rm expt}$.}
\label{t:vkl}
\begin{ruledtabular}
\begin{tabular}{l rrr rrr rrr r}
Ion/Atom & 
\multicolumn{1}{c}{State} &
\multicolumn{1}{c}{RHF} &
\multicolumn{1}{c}{$\Sigma^{(2)}$} &
\multicolumn{1}{c}{$\Sigma^{\infty}$} &
\multicolumn{1}{c}{Ladder} &
\multicolumn{1}{c}{Breit} &
\multicolumn{1}{c}{QED} &
\multicolumn{1}{c}{Final} &
\multicolumn{1}{c}{$\Delta$(\%)} &
\multicolumn{1}{c}{Expt.\footnotemark[1]} \\
\hline
Ba$^+$
& $6s_{1/2}$ & 75339 & 82379 & 80780 & -156 &  -4 & -45 & 80575 &-0.14 & 80687 \\
& $6p_{1/2}$ & 57265 & 61216 & 60571 & -128 & -27 &   3 & 60419 &-0.01 & 60425 \\ 
& $6p_{3/2}$ & 55873 & 59424 & 58847 & -118 &  -7 &   0 & 58722 &-0.02 & 58735 \\
& $5d_{3/2}$ & 68138 & 77444 & 76377 & -763 &  58 &  22 & 75694 &-0.16 & 75813 \\
& $5d_{5/2}$ & 67664 & 76500 & 75536 & -765 &  84 &  18 & 74873 &-0.19 & 75012 \\
Ra$^+$
& $7s_{1/2}$ & 75898 & 83864 & 82035 & -219 & -12 & -90 & 81714 &  0.16 & 81842 \\ 
& $7p_{1/2}$ & 56878 & 61432 & 60744 & -182 & -51 &   0 & 60511 &  0.03 & 60491 \\ 
& $7p_{3/2}$ & 52905 & 56278 & 55776 & -140 & -11 &  -3 & 55625 & -0.01 & 55633 \\ 
& $6d_{3/2}$ & 62355 & 71364 & 70294 & -620 &  72 &  42 & 69788 &  0.04 & 69758 \\ 
& $6d_{5/2}$ & 61592 & 69313 & 68563 & -643 &  92 &  33 & 68045 & -0.08 & 68099 \\ 
E120$^+$
& $8s_{1/2}$ & 83262 & 92195 & 90241 & -518 & -68 &-132 & 89523 & & \\
& $8p_{1/2}$ & 60040 & 66792 & 65448 & -378 &-125 & -16 & 64929 & & \\
& $8p_{3/2}$ & 49290 & 52744 & 52006 & -178 & -12 & -11 & 51805 & & \\
& $7d_{3/2}$ & 56610 & 66765 & 64815 & -590 &  68 &  61 & 64354 & & \\
& $7d_{5/2}$ & 56408 & 63526 & 62678 & -623 &  82 &  46 & 62183 & & \\
Fr
& $7s_{1/2}$ & 28767 & 34136 & 32924 & -136 &   5 & -47 & 32746 &  -0.3  &  32849 \\
& $7p_{1/2}$ & 18855 & 21004 & 20707 &  -76 & -14 &   0 & 20617 &   0    &  20612 \\
& $7p_{3/2}$ & 17655 & 19179 & 18971 &  -57 &   0 &  -1 & 18913 &  -0.06 &  18925 \\
& $6d_{3/2}$ & 13825 & 17190 & 16724 & -139 &  34 &  11 & 16630 &   0.07 &  16619 \\
& $6d_{5/2}$ & 13924 & 16849 & 16512 & -153 &  37 &   9 & 16405 &  -0.09 &  16419 \\
E119
& $8s_{1/2}$ & 33608 & 40489 & 39040 & -403 & -24 & -77 & 38536 & & \\
& $8p_{1/2}$ & 20130 & 23905 & 23473 & -184 & -47 &  -6 & 23236 & & \\
& $8p_{3/2}$ & 16672 & 18335 & 18114 &  -74 &  -1 &  -3 & 18036 & & \\
& $7d_{3/2}$ & 13477 & 17495 & 16807 & -149 &  34 &  19 & 16711 & & \\
& $7d_{5/2}$ & 13827 & 16899 & 16567 & -181 &  34 &  15 & 16435 & & \\
\end{tabular}
\footnotetext[1]{Ba$^+$ and Ra$^+$ data from Ref.~\cite{Moore},
  Fr data from Ref.~\cite{Fr-exp}.}
\end{ruledtabular}
\end{table*}

Results of calculations of the energies of the lowest $s$, $p$ and $d$
states of Ba$^+$, Ra$^+$, E120$^+$, Fr and E119 in different
approximations are presented in Table~\ref{t:vkl}. The RHF column
presents Hartree-Fock energies obtained by solving Eq.~(\ref{Brueck})
without $\hat \Sigma$, the $\hat \Sigma^{(2)}$ column presents Brueckner
energies obtained by solving Eq.~(\ref{Brueck}) with the second order correlation potential
$\hat \Sigma^{(2)}$. Note that since these energies are obtained by
solving the equations (\ref{Brueck}) rather than by calculating average
value of the correlation potential $\hat \Sigma^{(2)}$ as in (\ref{def-sigma}), they already
include one all-order effect, the iterations of $\hat \Sigma^{(2)}$. 
The $\hat \Sigma^{\infty}$ column presents Brueckner
energies obtained by solving Eq.~(\ref{Brueck}) with all-order
$\hat \Sigma^{\infty}$. The difference between this and previous
columns illustrate the importance of higher-order correlation effects
in $\hat \Sigma$, the screening of Coulomb interaction and
hole-particle interaction. 

The {\em Ladder} column presents contributions from ladder diagrams
given by (\ref{deltal}). We present these contributions separately for
the convenience of the discussion. We would like to emphasise the role
of ladder diagrams since it is the latest addition to our
all-order technique which has been tested before only for cesium and
thallium atoms~\cite{ladder}. If ladder diagrams are not included the
all-order correlation potential method developed in
Ref.~\cite{all-order} gives good accuracy for $s$ and $p$ states of
alkali atoms and their isoelectronic sequences. As it was
demonstrated in Ref.~\cite{ladder} adding ladder diagrams widens the
range of atomic systems for which the technique gives good
accuracy. The ladder diagram contributions do not affect much the $s$ and
$p$ states of alkali atoms while improve significantly the accuracy
for $d$ states. They also significantly improve the accuracy for such
complicated system as thallium atom~\cite{ladder}. As one can see from the
Table~\ref{t:vkl} ladder diagrams are important for all systems
considered in present paper leading to significant improvements of the
results. Breit and QED corrections are relatively small. However,
adding them generally leads to better agreement with experiment. The
data for Breit and QED corrections for Fr and E119 is in good
agreement with recent calculations by Thierfelder and
Schwerdtfeger~\cite{TS-QED}. Detailed discussion of the QED
corrections and comparision with other calculations presented in our
previous work~\cite{Dinh1}. 
Our final results for Ba$^+$, Ra$^+$ and Fr (see Table~\ref{t:vkl})
differ from the experimental data by small fraction of per cent only.

\begin{table*}
\caption{Removal energies (cm$^{-1}$) of Cs, Fr, E119 and
  E120$^+$; comparison with experiment~\cite{Moore,Fr-exp}, our ealier
  calculations~\cite{ladder,Dinh1} and the coupled cluster with single
  and double excitations (CCSD) calculations by
  Eliav {\em et al}~\cite{Eliav}.}
\label{t:final}
\begin{ruledtabular}
\begin{tabular}{l rrrr l rrrr}
Atom & 
\multicolumn{1}{c}{State} &
\multicolumn{1}{c}{This work\footnotemark[1]} &
\multicolumn{1}{c}{CCSD~\cite{Eliav}} &
\multicolumn{1}{c}{Expt.\footnotemark[2]} &
Atom/Ion & 
\multicolumn{1}{c}{State} &
\multicolumn{1}{c}{This work} &
\multicolumn{1}{c}{Ref.~\cite{Dinh1}} &
\multicolumn{1}{c}{CCSD~\cite{Eliav}} \\
\hline
Cs & $6s_{1/2}$ & 31384 & 31485 & 31407 & E119 & $8s_{1/2}$ & 38536 & 38852 & 38577 \\           
& $6p_{1/2}$ & 20185 & 20233 & 20229 &   & $8p_{1/2}$ & 23236 & 23272 & 22979 \\           
& $6p_{3/2}$ & 19632 & 19681 & 19675 &   & $8p_{3/2}$ & 18036 & 18053 & 18007 \\           
& $5d_{3/2}$ & 16932 & 19909 & 16908 &   & $7d_{3/2}$ & 16711 & & 16505 \\                       
& $5d_{5/2}$ & 16849 & 16809 & 16810 &   & $7d_{5/2}$ & 16435 & & 16297 \\                       

& $7s_{1/2}$ &       & 12886 & 12872 &   & $9s_{1/2}$ & 14061 & 14079 & 14050 \\           
& $7p_{1/2}$ &       &  9642 &  9641 &   & $9p_{1/2}$ & 10439 & 10415 & 10365 \\           
& $7p_{3/2}$ &       &  9462 &  9459 &   & $9p_{3/2}$ &  8882 & 8866 & 8855 \\             
& $6d_{3/2}$ &       &  8815 &  8818 &   & $8d_{3/2}$ &  8513 & & 8455 \\                       
& $6d_{5/2}$ &       &  8772 &  8775 &   & $8d_{5/2}$ &  8399 & & 8338 \\                       
& $8s_{1/2}$ &       &  7082 &  7090 &   & $10s_{1/2}$&  7521 & 7536 & 7519 \\             
& $8p_{1/2}$ &       &  5689 &  5698 &   & $10p_{1/2}$&  6024 & 6018 & 5997 \\             
& $8p_{3/2}$ &       &  5606 &  5615 &   & $10p_{3/2}$&  5334 & 5328 & 5320 \\             
& $7d_{3/2}$ &       &  5354 &  5359 &   & $9d_{3/2}$ &  5177 & & 5154 \\                       
& $7d_{5/2}$ &       &  5333 &  5338 &   & $9d_{5/2}$ &  5118 & & 5092 \\                       

Fr & $7s_{1/2}$ & 32746 & 32930 & 32849 & E120$^+$ & $8s_{1/2}$ & 89523 & 89931 & \\                 
& $7p_{1/2}$ & 20617 & 20597 & 20612 &         & $8p_{1/2}$ & 64929 & 65080 & \\                 
& $7p_{3/2}$ & 18913 & 18918 & 18925 &         & $8p_{3/2}$ & 51805 & 51874 & \\                 
& $6d_{3/2}$ & 16630 & 16527 & 16619 &         & $7d_{3/2}$ & 64354 & & \\                       
& $6d_{5/2}$ & 16405 & 16339 & 16419 &         & $7d_{5/2}$ & 62183 & & \\                       
& $8s_{1/2}$ & 13075 & 13131 & 13116 &         & $9s_{1/2}$ & 40085 & 40110 & \\                 
& $8p_{1/2}$ &  9730 &  9732 &  9736 &         & $9p_{1/2}$ & 32618 & 32604 & \\                 
& $8p_{3/2}$ &  9184 &  9190 &  9191 &         & $9p_{3/2}$ & 27978 & 27951 & \\                 
& $7d_{3/2}$ &  8584 &  8597 &  8604 &         & $8d_{3/2}$ & 31489 & & \\                       
& $7d_{5/2}$ &  8490 &  8507 &  8516 &         & $8d_{5/2}$ & 30727 & & \\                       
& $9s_{1/2}$ &  7160 &  7184 &  7178 &         & $10s_{1/2}$& 23307 & 23357 & \\                 
& $9p_{1/2}$ &  5726 &  5738 &       &         & $10p_{1/2}$& 19887 & 19926 & \\                 
& $9p_{3/2}$ &  5477 &  5493 &       &         & $10p_{3/2}$& 17664 & 17678 & \\                 
& $8d_{3/2}$ &  5209 &  5243 &  5248 &         & $9d_{3/2}$ & 19293 & & \\                       
& $8d_{5/2}$ &  5162 &  5198 &  5203 &         & $9d_{5/2}$ & 18921 & & \\                       
\end{tabular}
\footnotetext[1]{Results for Cs are taken from Ref.~\cite{ladder}.}
\footnotetext[1]{Cs data from Ref.~\cite{Moore},
  Fr data from Ref.~\cite{Fr-exp}.}
\end{ruledtabular}
\end{table*}

Final results for Fr and superheavy elements E119 and E120$^+$ are presented
in Table~\ref{t:final} together with the results of our previous
calculations for Cs~\cite{ladder}, E119 and E120$^+$~\cite{Dinh1}, the
results of the coupled cluster calculations by Eliav {\em et al} for
Cs, Fr and E119 and experimental data for Cs and Fr. 
Judging by the data in the tables table we believe that the accuracy of
the calculated energies is on the level of 0.2\%. 
%Table~\ref{t:final} presents also results of our previous
%calculations~\cite{Dinh1}. 

There are two important differences between present calculations and those
of Ref.~\cite{Dinh1}. Ladder diagrams were not included in
\cite{Dinh1}. On the other hand the {\em ab initio} results for superheavy
elements were corrected in \cite{Dinh1} by extrapolating of the theoretical
error from lighter elements. This extrapolation assumes similar
electron structure of the elements. In contrast, present calculations
are pure {\em ab initio} calculations with no fitting or
extrapolating. Calculation of ladder diagrams reveal some small
differences in electron structure of superheavy elements and their
lighter analogs. Indeed, the contribution of ladder diagrams tends
to be larger for superheavy elements. This is consistent with larger
removal energies. Larger removal energies means that valence
electron is closer to the core, therefore its residual Coulomb
interaction with the core described by ladder diagrams should be
larger as well. Note that the difference between our present and
previous~\cite{Dinh1} results is sometimes larger than 0.2\% (the
accuracy of present calculations). This is particularly true for the
ground state energies. We believe that the accuracy of present
calculations is better that in Ref.~\cite{Dinh1} since they are pure
{\em ab initio} calculations with no fitting and no extrapolation and
they do take into account small differences in electron structure
between superheavy elements and their lighter analogs. 

Table~\ref{t:final} shows very good agreement between present results
and the results of the coupled cluster calculations of
Ref.~\cite{Eliav}. The agreement is better than with our previous 
calculation~\cite{Dinh1}.
In the end, both methods, the method of present work and the CCSD 
method used in Ref.~\cite{Eliav} demonstrate very similar levels of accuracy. 
%It is interesting to note that in 
%the cases of large differences with our previous results our new
%results are closer to those of Ref.~\cite{Eliav}.

Table~\ref{t:final} presents also the energies of $d$ states of E119
and E120$^+$. Correlations for $d$ states are usually larger than for
$s$ and $p$ states and accuracy of calculations is lower. That was
part of the reason why these states were not considered
before. However, as it is evident from the data in Table~\ref{t:vkl},
the inclusion of ladder diagrams leads to practically the same
accuracy for $d$ states as for $s$ and $p$ states. Therefore, we
include the results for the energies of $d$ states of superheavy
elements in Table~\ref{t:final}.

\section{Conclusion}

The result of this paper is twofold. First, we demonstrate that
supplementing previously developed all-order correlation potential
method with ladder diagrams leads to significant improvement in
accuracy of calculations not only for alkali atoms but also for their 
isoelectronic ions. Second, we apply the developed technique to
perform very accurate calculations of the spectra of superheavy
elements E119 and E120$^+$.

\acknowledgments

The author is grateful to V. V. Flambaum and M. G. Kozlov for useful
discussions. 
The work was supported in part by the Australian Research Council.

\end{document}